\documentclass[preprint,10pt]{aastex}
\usepackage{emulateapj5}

\def\cm2{\,{\rm cm}^{-2} }

\def\hmpc{\,h^{-1}\,{\rm Mpc} }

\def\kpc{\, {\rm kpc} }

\def\kmsmpc{\,{\rm km\, s^{-1}\, Mpc^{-1}} }

\def\ergcms{\,{\rm erg\, cm^{-2}\, s^{-1}} }

\def\ergs{\,{\rm erg\, s^{-1}}}
\def\ergsh2{\,{\rm erg\, s^{-1}}, h=\frac{1}{2} }

\def\Jy{\,{\rm Jy} }

\def\mJy{\,{\rm mJy} }

\def\muK{\,\mu{\rm K} }

\def\GHz{\,{\rm GHz}}
\def\Hz{\,{\rm Hz}}

\def\etal{{\it et al. }}

\begin{document}


\title{A Measurement of ${\rm H_0}$ from the Sunyaev-Zeldovich Effect}

\author{Brian S. Mason \altaffilmark{1,2}, Steven T. Myers
\altaffilmark{1,3}, and A.C.S. Readhead\altaffilmark{4}}

\altaffiltext{1}{University of Pennsylvania  209 S. 33rd St. Philadelphia, PA 19104-6396}

\altaffiltext{2}{ Current Address:  California Institute of Technology, 105-24,  Pasadena, CA 91125}

\altaffiltext{3}{ Current Address: NRAO, Socorro, NM 87801}

\altaffiltext{4}{ California Institute of Technology, 105-24,  Pasadena, CA 91125}

\begin{abstract}
We present a determination of the Hubble constant, ${\rm H_0}$, from
measurements of the Sunyaev-Zeldovich Effect (SZE) in an
orientation-unbaised sample of seven $z< 0.1$ galaxy clusters.  With
improved X-ray models and a more accurate $32 \, \GHz$ calibration, we
obtain ${\rm H_0} = 64^{+14}_{-11} \pm 14_{sys} \kmsmpc$ for a
standard CDM cosmology, or ${\rm H_0} = 66^{+14}_{-11} \pm 15_{sys}
\kmsmpc$ for a flat $\Lambda{\rm CDM}$ cosmology.  In combination with
X-ray cluster measurements and the BBN value for $\Omega_B$, we find
$\Omega_M = 0.32 \pm 0.05$.

\end{abstract}


\section{Introduction}
  
The Sunyaev-Zeldovich effect (SZE) is a spectral distortion in the
Cosmic Microwave Background (CMB) due to inverse Compton scattering of
the CMB photons off hot electrons; at radio frequencies this
distortion is manifested as a fractional decrement in intensity of
order $10^{-4}$. For over two decades it has been known
\citep{Silk_and_White_1978, Cavaliere_et_al_1979} that the combination
of X-ray and SZE observations of rich galaxy clusters, under the
assumption of spherical symmetry, yields a direct measurement of the
cosmic distance scale.  Only in the past few years have reliable and
accurate applications of this method become possible
\citep[e.g.][]{Birkinshaw_et_al_1991,Herbig_et_al_1995,
Carlstrom_Joy_and_Grego_1996,Grainge_et_al_1996,Holzapfel_et_al_1997}.
Due to the assumption of spherical symmetry, selection biases have
been a great concern.  To address this \citet{Myers_et_al_1997}
defined an X-ray flux-limited sample of 11 $z < 0.1$ clusters and
began a campaign to measure distances to these clusters.  This yields
an orientation-unbiased sample, and departures from spherical symmetry
in individual clusters will average out in the determination of ${\rm
H_0}$ from the sample as a whole.  Myers \etal obtain a Hubble
Constant of $54 \pm 14 \, \kmsmpc$ from observations of four clusters.
The accuracy of this result is limited by a 7\% radio calibration
uncertainty and estimated 15-30\% X-ray model uncertainties for each
cluster.

In this Letter we present an improved measurement of ${\rm H_0}$ based
on the Herbig \etal and Myers \etal results on Coma, A478, A2142, and
A2256, plus observations of three new clusters (A399, A401, A1651)
from their complete sample.  The results we present incorporate more
accurate X-ray models \citep[][hereafter,MM2000]{Mason_and_Myers_2000}
and a more accurate radio calibration \citep{Mason_et_al_1999}.  We
also calculate the effect of intrinsic CMB anisotropies on our result
and find them to be a limiting factor for a sample of this size.
Throughout this paper we use ${\rm H_0} = 100 \, h \kmsmpc$, and
consider two cosmologies: SCDM ($\Omega_M=1, \Omega_{\Lambda}=0$), and
$\Lambda {\rm CDM}$ ($\Omega_M=0.3, \Omega_{\Lambda}=0.7$).  Unless
otherwise stated, we assume the SCDM cosmology.

\section{The Sunyaev-Zeldovich Effect, Cluster Sample, and X-Ray Models}

\subsection{The Sunyaev-Zeldovich Effect}

The observed fractional change in antenna temperature induced by a
cloud of electrons with a beam-averaged Compton $y-$parameter
$y_{obs}$ is \citep{Sunyaev_and_Zeldovich_1980}
\begin{equation}
\label{eq:dt_ant_sze}
\frac{\Delta  T_{obs}}{ T_{cmb}}  =  \frac{x^2 e^x}{(e^x - 1)^2} 
\left[ x \coth (x/2) - 4 \right] \chi_{rel}^{-1} \times \, y_{obs}.
\end{equation}
Here $x=h\nu/kT_{cmb}$, and $\chi_{rel}^{-1}$ is a relativistic
correction factor which we compute using the analytic expression of
\citet{Sazonov_and_Sunyaev_1998}.  We assume zero peculiar velocity,
which introduces a $\sim 5\%$ uncertainty but will average out over
the sample.  Along a given line of sight, the Compton $y-$parameter is
\begin{equation}
y = \frac{kT_e}{m_ec^2} \, \tau
\end{equation}
where $\tau$ is the inverse Compton optical depth.
At $32.0 \, \GHz \, (x=0.563)$, Eq.~\ref{eq:dt_ant_sze} takes the form
\begin{equation}
\frac{\Delta T_{obs}}{T_{cmb}} = -1.897 \times y_{obs} \, \chi_{rel}^{-1}.
\end{equation}
We  follow the procedure of Myers  \etal  in correcting the
observed Compton $y-$parameters for relativistic effects rather than
the models.

Given a model for the cluster density and temperature profiles from
X-ray observations, we predict the SZE decrement in terms of the
Compton $y-$parameter averaged over the telescope beam and switching
patterns:
\begin{equation}
y_{pred} =  \frac{1}{\Omega_{beam}}  \, \int \, d\Omega \, y_{model}(\widehat{\Omega}) \, R_{N,sw}(\widehat{\Omega}).
\label{eq:ysw}
\end{equation}
Here $\Omega_{beam}$ is the solid angle of the telescope beam,
$y_{model}$ is the predicted $y$ along a given line of sight, and
$R_{N,sw}(\widehat{\Omega})$ is the normalized beam response,
including the effects of switching. It is easily shown that
$y_{pred}$, as determined from X-ray observations and Eq.~\ref{eq:ysw},
is proportional to $h^{-1/2}$, so that if $y_{obs} = q$ and $y_{pred}
= p \, h^{-1/2}$, the Hubble constant is given by
\begin{equation}
\label{eq:pq}
h = \left( \frac{p}{q} \right)^2.
\end{equation}
Note that the $\tau_{sw}$ values presented in MM2000 are simply the
model inverse Compton optical depths convolved with the telescope beam
switching pattern as per Eq.~\ref{eq:ysw}.  For an isothermal cluster,
$y_{pred} = \tau_{sw} \, kT_e/m_e c^2$.
   
\subsection{Cluster Sample}

In MM2000 we define a larger, X-ray flux-limited cluster sample which
expands upon that presented by Myers \etal.  This is a 90\%
volume-complete sample selected from the XBAC catalog
\citep{Ebeling_et_al_1996} with $z <0.1$ and $F_X > 1.0 \times
10^{-11} \ergcms$.  At $z=0.1$, the volume-completeness criterion
corresponds to $L_X > 1.13 \times 10^{44} h^{-2} \ergs$ (0.1 - 2.4
keV).  The resulting set of 31 clusters contains the Myers \etal
sample.  While the 7 clusters for which we present measurements here
are all members of the smaller Myers \etal sample, this work is part
of an ongoing project to survey distances to the objects in our
expanded sample.  MM2000 discusses in detail the completeness of this
XBAC-derived catalog.

\subsection{X-Ray Models}

MM2000 also presents X-ray models for the 22 clusters in our sample
which had public ROSAT PSPC data as of May 1999.  The primary focus of
this analysis was to quantify the uncertainty in the beam-convolved
inverse Compton optical depth, $\tau_{sw}$, a quantity which we find
is robustly constrained by the ROSAT data.  We adopt the
\citet{Markevitch_et_al_1998} measurements of the cluster gas
temperature $T_e$, which--- except for the Coma cluster--- we assume
to be isothermal.  For the Coma cluster, following
\citet{Hughes_et_al_1988a}, we adopt a hybrid model which is
isothermal inside a radius of $500 \, h^{-1} \kpc$, outside of which
the temperature follows the density profile with an adiabatic index of
$\gamma = 1.5$.  For A478 and A2142, we use the fits which have been
corrected for the spectral bias induced by the cooling flow emission.
Table~\ref{tbl:modelpars} summarizes our cluster models and the
resulting beam-averaged Compton $y-$values, $y_{pred}$.  In
\S~\ref{sec:interp} we discuss the impact on ${\rm H_0}$ of our
choices for the cluster models.

\begin{table*}[t]
\begin{center}
\caption{Cluster Model Parameters}
\label{tbl:modelpars}
\begin{tabular}{ l l l l l l l }
\tableline\tableline
Cluster &  $z$ & $\theta_o$ & $\beta$ & $n_{eo}$ & $k T_{e}$ &
$\tau_{sw}$ \\ 
        &      &  ($\,'\,$) &         & ($10^{-3} \, h^{1/2} \, {\rm
cm^{-3}}$) & (keV) & ($10^{-3} \, h^{-1/2}$)\\ \tableline
A399  &  $0.0715$  & $4.33 \pm 0.45$     & $0.742 \pm 0.042$      &
$3.23 \pm 0.18$  & $7.0 \pm 0.2$  & $2.50 \pm 0.13$    \\
A401  & $0.0748 $    &  $2.26 \pm0.41$  &  $0.636 \pm 0.047 $    &
$7.90 \pm 0.81$ &  $8.0 \pm 0.2$  & $3.17 \pm 0.18 $  \\
A478  & $0.0900$ & $1.00 \pm 0.15$ & $0.638 \pm 0.014$ & $27.81 \pm
9.7$  &  $8.4 \pm 0.7 $  & $3.68 \pm 0.15 $   \\
A1651 &  $0.0825$ & $2.16 \pm 0.36$ & $0.712 \pm 0.036$ & $7.14 \pm
3.20$  &  $6.1 \pm 0.2 $  & $2.44 \pm 0.11$   \\
Coma  & $0.0232$ & $9.32 \pm 0.10$  &  $0.670 \pm 0.003$ &  $4.52 \pm
0.04$ & $9.1 \pm 0.4$ & $2.76 \pm 0.16$  \\
A2142 & $0.0899$ &   $1.60 \pm 0.12$ & $0.635 \pm 0.012$ &  $14.95 \pm
1.0$ &  $9.7 \pm 0.8$ & $4.28 \pm 0.18$  \\
A2256 & $0.0601$  &  $5.49 \pm 0.21$ &  $0.847 \pm 0.024$ & $4.08 \pm
0.08$  &  $6.6 \pm 0.2$   &  $3.19 \pm 0.16$ \\ \tableline
\multicolumn{7}{l}{NOTE: $\tau_{sw}$, $\theta_0$, $\beta$, and $n_{eo}$ are from
MM2000; $T_e$ values are from \citet{Markevitch_et_al_1998}; 
} \\
\multicolumn{7}{l}{redshifts are from \citet{Struble_and_Rood_1991_redshifts}. Uncertainties are $1\sigma$.}
\end{tabular}
\end{center}
\end{table*}

\section{SZE Observations}

In this section we report on observations of 7 clusters from our
sample.  Two of these, A399 and A401, are in close proximity.  The
effects of this are included in our model predictions (see
\S~\ref{sec:interp}).

%
%

\subsection{New Observations}

For the observations reported here, the OVRO 5.5-meter telescope (5-m)
was outfitted with a HEMT receiver having a center frequency of 32 GHz
and a bandwidth of 6.5 GHz.  The HEMT input is Dicke switched every
millisecond between two ambient-temperature corrugated feeds which
give rise to two 7'.35 FWHM Gaussian beams [each having a main-beam
volume of $\Omega_{beam} = (5.21 \pm 0.03) \times 10^{-6} \, {\rm
Sr}$] separated by 22'.16 in azimuth.  To remove systematic effects
due primarily to atmosphere and ground, we employ a
triple-differencing technique. Two levels of differencing are provided
by the Dicke switching between the target (ON) position and a
reference (REF) position, and and by nodding the telescope in azimuth
at a rate of $\sim 0.1 \Hz$.  The third level is provided by observing
blank leading (LEAD) and trailing (TRAIL) fields over the same range
of azimuth and elevation as the MAIN field, so that ground-based
signals are cancelled in the difference.  The cluster signal is given
by the MAIN field signal minus the average of the LEAD and TRAIL
signals, or MLT.  The LEAD-TRAIL field difference (LTD) provides a
diagnosis of possible residual signals due, {\it e.g.}, to
unsubtracted ground emission or intrinsic CMB fluctuations.

Flux density calibration is accomplished with reference to Cas A using
an epoch 1998 flux of $194 \pm 4 \Jy$ \citep{Mason_et_al_1999}, with
the secular variation of Cas A modeled as per
\citet{Baars_et_al_1977}.  Data are edited by automatic filtering
algorithms that use the scatter of the data and their standard
deviations as rejection criteria.  More discussion of our observing
strategy and data filters is given in \citet{Readhead_et_al_1989} and
\citet{Myers_et_al_1997}; more details on the instrument are in
\citet{Leitch_et_al_2000}.

\subsubsection{Discrete Source Removal}
\label{sec:ptsrc}

The brightest contaminating discrete sources in A399 and A1651,
selected from 87GB \citep{Gregory_and_Condon_1991}, were monitored at
18.5 GHz with the OVRO 40-meter telescope in November and December of
1996, concurrent with cluster SZE observations.  In the Fall of 1997
we used the OVRO 40-m telescope to survey all NVSS
\citep{Condon_et_al_1998} sources in these fields and those for A401
which, assuming a flat spectrum, would give a peak signal at 32 GHz $>
16 \muK$. We also observed {\it all} sources within $12'$ of any of
the field centers and having 1.4 GHz flux densities $> 50 \mJy$.
Sources which showed indications of variability were extrapolated to
32 GHz assuming a flat spectrum; otherwise the two-point spectral
index between 1.4 and 32 GHz was used.  The overall corrections in
$\muK$ are given for each cluster in \S~\ref{sec:newclusters}.

\subsubsection{A399, A401, and A1651}
\label{sec:newclusters}

Observations of A399 were taken from October 1996 through March 1997.
For these observations the LEAD and TRAIL fields were separated from
the MAIN by 26 minutes of Right Ascenscion.  After statistical filters
and weighting of the data, we acquired 40 hours of total integration
time including LEAD and TRAIL fields, and time spent on the REF beams.
To account for source contamination $24 \pm 5\muK$ was subtracted from
the observed signal.  We then determine a decrement of $\Delta T_{obs}
= -164 \pm 21 \muK$ (MLT).  The LEAD-TRAIL difference is $\Delta
T_{\rm LTD} = 15 \pm 21 \muK$.  The observed $\Delta T$ corresponds to
$y_{obs} = (3.28 \pm 0.42)\times 10^{-5}$, including the relativistic
correction $\chi_{rel} = 1.027$.

A1651 was also observed during this period, with LEAD and TRAIL field
separations of $20^{\rm min} \, 30^{\rm sec}$ from the MAIN field; the
effective total integration time was 25 hours.  We subtract $5 \pm 1
\, \muK$ from the observed decrement to correct for discrete sources.
These are mostly in the REF beam of the LEAD field. With the source
correction we determine a decrement $\Delta T_{obs} = -247 \pm 30
\muK$ (MLT).  The LTD is $\Delta T_{\rm LTD} = -92 \pm 36 \muK$.  The
Compton-$y$ parameter is $y_{obs} = (4.88 \pm 0.59) \times 10^{-5}$
including the relativistic correction $\chi_{rel} = 1.023$.

A401 was observed from October 1997 to March 1998, giving an effective
total integration time of 43 hours.  The LEAD and TRAIL field
separations were $16^{\rm min} \, 36^{\rm sec}$.  Discrete sources
contribute $16 \pm 11 \, \muK$.  The source-corrected decrement is
$\Delta T_{obs} = -338 \pm 20 \muK$ (MLT); the LTD is $\Delta T_{\rm
LTD} = 133 \pm 24 \muK$.  This gives (with $\chi_{rel} = 1.031$)
$y_{obs} = (6.33 \pm 0.43) \times 10^{-5}$.

The statistically significant LTD's for A401 and A1651 are discussed
in \S~\ref{sec:interp}.

\subsection{ Coma Cluster }

\citet{Herbig_et_al_1995} used the OVRO 5-m telescope to determine the SZE
decrement towards the Coma cluster (Abell 1656) during the observing
seasons of 1992 and 1993, and found a Rayleigh-Jeans temperature
decrement of $\Delta T_{obs} = -302 \pm 48 \muK$.  These data were
calibrated relative to DR21 assuming $S_{DR21} = 18.24 \pm 0.55 \Jy$,
and referenced to the telescope main beam assuming $\Omega_{Beam} =
(5.16 \pm 0.15) \, \times 10^{-6} \, {\rm Sr}$.  From the Mason et
al. flux density scale and $\chi_{rel} = 1.035$, we determine a
beam-averaged Compton $y-$parameter of $y_{sw} = (6.38 \pm 1.01)
\times 10^{-5}$.

\subsection{ A478, A2142, and A2256}

A478, A2142, and A2256 were observed by \citet{Myers_et_al_1997} from
July 1993 to March 1994 resulting in decrements of $-375 \pm 28 , -437
\pm 25 , \,$ and $-243 \pm 29 \muK \,$ respectively; these temperature
are referred to the main beam, and have been corrected for
point-source contamination.  These data were calibrated using a
brightness temperature of $T_J = 144 \pm 8\,$ K for Jupiter.
After including the correction for the main beam area [Myers  \etal  use
$(5.12 \pm 0.14) \times 10^{-6} \, {\rm Sr}$], we find an overall
correction of $f = 1.037$ for the Myers  \etal  data.  The stated
decrements yield switched, beam-averaged Compton-$y$ parameters of
$(7.25 \pm 0.54)\times 10^{-5} \, , (8.44 \pm 0.48) \times 10^{-5} ,
\,$ and $(4.70 \pm 0.56) \times 10^{-5}$, respectively.  We apply
relativistic corrections ($\chi_{rel} = 1.033 \,$ , 1.037 , and 1.026)
together with our calibration correction, and find Compton $y-$values
of $(7.77 \pm 0.58) \times 10^{-5}$ , $(9.10 \pm 0.52) \times 10^{-5}$
, and $\, (5.00 \pm 0.60) \times 10^{-5}$.

\section{Interpretation}
\label{sec:interp}

For SZE observations at our frequencies and angular scales intrinsic
CMB anisotropies are a significant source of uncertainty.  We use the
RING5M measurements \citep{Leitch_et_al_2000} to determine the impact
of this on our ${\rm H_0}$ results.  In measurements taken with the
OVRO 5-m telescope at 32 GHz, Leitch \etal determined an RMS
temperature difference on $22'$ scales of $\delta T_{22'} =
79^{+11}_{-10} \, \mu {\rm K}$.  In order to determine the degree to
which parallactic angle averaging during a track on a cluster reduces
this, we generated $10^3$ realizations of $4 \, {\rm deg^2}$ patches
of sky from a $\Lambda$CDM power spectrum.  Each realization was
convolved with the 5-m main beam and the switching pattern
characteristic of a typical SZE observation.  We find (including the
LEAD/TRAIL differencing) a residual CMB signal of $81.6\% \, \times
\delta T_{22'}$, or $64 \, \muK$ ($\sigma_y = 1.24 \times 10^{-5}$)
assuming the RING5M power level.  The RMS of LTD predicted by this
analysis is $90 \, \muK$, close to the observed RMS of $79 \, \muK$ in
our 7 clusters.  On this basis the non-zero LTD's seen in A1651 and
A401 are not unexpected.

While this noise level is not small compared to the signal in fainter
clusters such as A399, we leave these clusters in our sample to avoid
an orientation bias and perform a Maximum Likelihood analysis to
determine ${\rm H_0}$.  Assume that we have $N$ observed Compton-$y$
paramaters $q_i$, and $N$ predicted $y$-parameters (as per
Eq.~\ref{eq:pq}) $p_i$ determined from X-ray observations.  The SZE
observations are corrupted with random Gaussian noise $\sigma_{q,i}$
(including thermal noise and point-source subtraction uncertainties)
and the predictions are corrupted with random Gaussian noise
$\sigma_{p,i}$; the RMS residual CMB signal is $\sigma_{cmb} = 64 \,
\muK$.  In addition there are $N$ ``true'' predictions $\widehat{p_i}$
which are the predictions which we would have in the absence of errors
in the X-ray models.  Then we can determine $h$ by minimizing
\begin{equation}
\label{eq:chi2}
\chi^2 = \sum_{i=1}^N \frac{(q_i - h^{-1/2}
 \widehat{p_i})^2}{\sigma_{q,i}^2 + \sigma_{cmb}^2 } + \frac{(p_i - \widehat{p_i})^2}{\sigma_{p,i}^2},
\end{equation}
where $\chi^2 \equiv -2 \, ln \, L$.  The unconstrained
$\widehat{p_i}$ must also be minimized in this process, but these can
be projected out analytically.  Setting the derivatives of $\chi^2$
with respect to the $\widehat{p_i}$ equal to zero, we find
\begin{equation}
\widehat{p_i} = \frac{ q_i + f_i h^{1/2} p_i}{h^{-1/2} + f_i h^{1/2} }
\end{equation}
with $f_i = (\sigma_{q,i}^2 + \sigma_{cmb}^2)/\sigma_{p,i}^2$.  This
leaves a 1-dimensional parameter space.  The value of $h$ which
minimizes Eq.~\ref{eq:chi2} is the most probable value; two-sided 68\%
confidence intervals are determined about this value by integrating
the likelihood function $L$.  We impose the prior $0.01 < h < 5$; our
upper bound on ${\rm H_0}$ for A399 is slightly affected by our choice
of prior, but other results are not.  Monte Carlo tests with the
estimator of Eq.~\ref{eq:chi2} show that within the range of power
levels allowed by RING5M, our result is not affected by the
uncertainty in $\sigma_{cmb}$.

Table~\ref{tbl:little_ho} shows the measured and predicted values of
the Compton $y-$parameter for our 7 clusters along with the
Maximum-Likelihood values for and 68\% confidence limits on ${\rm
H_0}$.  A399 and A401 are in close spatial proximity with the result
that the observed decrement on either is reduced somewhat due to
signal from the other in the reference beam; this is accounted for in
the stated values of $y_{pred}$.  Averaged over a track, we find a
decrease in the predicted $y$ for A399 of $(0.15 \pm 0.05) \times
10^{-5} \, h^{-1/2}$ and for A401 of $(0.13 \pm 0.05) \times 10^{-5}
\, h^{-1/2}$.  For an SCDM cosmology, the sample average is
$64^{+14}_{-11} \kmsmpc$.  The {\it scatter} of the ${\rm H_0}$ values
gives an error in the mean of only $8 \kmsmpc$ , which is less than
that given by our Maximum Likelihood analysis.  The minimum
of $\chi^2$ corresponds to $\chi_{\nu}^2 = 0.32$ for $\nu = 6$; there
is a $\sim 5\%$ chance of obtaining a $\chi_{\nu}^2$ value this low by
chance for six degrees of freedom.  We therefore think it likely that
the scatter in these 7 clusters is fortuitously small.  Since our
observing and modelling errors have been well quantified, the Maximum
Likelihood method gives a more reliable estimate of the uncertainty
than the data scatter for a small sample such as ours.  Our result is
not sensitive to the model we choose for the cluster temperature
profile: adopting hybrid models for all 5 clusters reduces the sample
average by only 4\%.  Although more extreme models for the temperature
profiles would have a larger effect on our result, such models are not
motivated by current X-ray analyses
\citep{Irwin_et_al_1999,White_2000}.  For a ${\rm \Lambda CDM}$
cosmology, the sample average is $ 66^{+14}_{-11} \kmsmpc$.  The
calibration uncertainties are $3\%$ (radio) and $8\%$ (X-ray), and we
estimate a $10\%$ uncertainty in the SZE predictions due to the
possibility of substructure and non-isothermality in the ICM.
Altogether we have a systematic error budget of $ \pm 14_{sys}
\kmsmpc$, or $\pm 15_{sys} \kmsmpc$ for ${\rm \Lambda CDM}$.

The increase in ${\rm H_0}$ relative to the Myers \etal result is due
primarily to differences in the updated X-ray models.  As a check on
our ROSAT-derived density models, we have compared the baryonic masses
that we obtain inside $500 \, h^{-} \kpc$ to those obtained by
\citet{Mohr_et_al_1999} for 20 of the 22 clusters in common to our
analyses: we find that the mean mass ratio is equal to unity at better
than $1\%$ accuracy.  See MM2000 for more discussion of this analysis.
Also the $T_e$ values we have used tend to be higher than the Myers
\etal values due to the correction for the spectral bias of the
cooling flows.  Using the Myers \etal temperatures increases the
fractional scatter in the ${\rm H_0}$ values by over a factor of two
and decreases the sample mean to $55 \kmsmpc$.  The fact that the
uncertainty in our result is comparable to that of Myers \etal--- in
spite of improved X-ray models, better calibration, and 3 more
clusters--- is due to our inclusion of intrinsic anisotropy in the
analysis.

\begin{table*}[t]
\centering
\caption{${\rm H_0}$ results on 5 Clusters}
\begin{tabular}{l l l c}
\tableline\tableline
Cluster & $y_{obs} $ & $y_{pred} $  & ${\rm
H_0}$ \\ 
 &$(10^{-5})$ & $(10^{-5} h^{-1/2})$ & $(\kmsmpc)$ \\ \tableline
A399 & $3.24 \pm 0.41$ & $3.27 \pm 0.20$      & $102^{+116}_{-53}$ \\
A401 & $6.93 \pm 0.46$ & $4.82 \pm 0.32$      & $48^{+28}_{-16} $ \\
A478 & $7.77 \pm 0.58$ & $6.05 \pm 0.54$      & $61^{+33}_{-20}$ \\
A1651 & $4.88 \pm 0.59$ & $2.92 \pm 0.17$     & $36^{+35}_{-15} $ \\
Coma & $6.38 \pm 1.01$ & $5.00 \pm 0.38$     & $62^{+49}_{-24}$ \\
A2142 & $9.10\pm0.52$   & $8.12  \pm  0.74 $  & $79^{+34}_{-24}$ \\
A2256 & $5.00 \pm 0.60$ & $ 4.11  \pm  0.26$  & $67^{+62}_{-28}$ \\ 
SAMPLE & --- & ---  & ${\bf 64^{+14}_{-11}}$ \\ \tableline
\multicolumn{4}{l}{NOTE: Uncertainties are $1\sigma$ random errors only.}\\
\end{tabular}
\label{tbl:little_ho}
\end{table*}

\section{Conclusions}

We have presented a determination of ${\rm H_0}$ resulting from
measuring the SZE in 5 clusters from an unbiased, low-redshift sample.
For an SCDM cosmology we find ${\rm H_0} = 64^{+14}_{-11} \pm 14_{sys}
\kmsmpc$, while for a ${\rm \Lambda CDM}$ cosmology we obtain ${\rm
H_0} = 66^{+14}_{-11} \pm 15_{sys} \kmsmpc$.  This result is in good
agreement with other recent measurements.  \citet{Madore_et_al_1999},
using Cepheid variables as distance indicators, find $72 \pm 5 \pm
7_{sys} \,\kmsmpc$.  Gravitational lens time delays give results
consistent with these \citep[e.g.][]{Fassnacht_et_al_1999}.

Our value of ${\rm H_0}$ implies an age for the universe ranging from
$(10.2 \pm 3.2) \times 10^9\,$ years for an SCDM universe to $(14.2
\pm 4.5)\times 10^9 \, $ years for ${\rm \Lambda CDM}$.  These are
both consistent with recent age determinations from main-sequence
fitting, which give ages of $(> 12 \pm 1)\times 10^9 \,$ years
\citep{Chaboyer_et_al_1998}.  In the X-ray cluster analysis of MM2000
we find a mean ICM mass fraction $f_{ICM} \equiv M_{bary}/M_{tot}=
(7.02 \pm 0.28 ) \, h^{-3/2}\, \times 10^{-2}$ (within $R_{500} \sim 1
\hmpc$) in a sample of 22 nearbly clusters.  We combine this result
with the Big Bang Nucleosynthesis constraint $\Omega_B h^2 = 0.019 \pm
0.001\,$ \citep{Burles_and_Tytler_1998} and our value for ${\rm H_0}$
to find a total matter density $\Omega_M = 0.32 \pm 0.05$, arguing
against SCDM cosmologies.

\acknowledgements

We are grateful to Russ Keeney for his many years of hard work on the
5-m and 40-m telescopes at OVRO, and to Erik Leitch for writing the
software that was used to analyze the OVRO data. We thank Jonathan
Sievers for assistance with the intrinsic CMB simulations and the
1997/1998 OVRO observations.  We received suport from NSF grants AST
91-19847 and AST 94-19279.  For part of the duration of this work BSM
was supported by the Zacheus Daniels fund at the University of
Pennsylvania.  STM was supported by an Alfred R. Sloan Fellowship at
the University of Pennsylvania.  The National Radio Astronomy
Observatory is a facility of the National Science Foundation operated
under cooperative agreement by Associated Universities, Inc.

\bibliographystyle{apj}

\end{document}